%% file: Al_Moulla_et_al_2022.tex
\documentclass{aa}
\usepackage[utf8]{inputenc}
\usepackage[T1]{fontenc}

\usepackage[english]{babel}
\usepackage{txfonts}
\usepackage{charter}
\usepackage{balance}
\usepackage{hyperref}
    \hypersetup{colorlinks=true,linkcolor=blue,citecolor=blue,urlcolor=blue}
    \makeatletter
    \renewcommand*\aa@pageof{, page \thepage{} of \pageref*{LastPage}}
    \makeatother

\usepackage{siunitx}
\usepackage{bm}

\usepackage{booktabs}
\usepackage[flushleft,para]{threeparttable}
\usepackage{multirow}

\usepackage[nowarn]{glossaries}
    \makeglossaries
    \loadglsentries{Glossary}

\begin{document}

   %Title
   \title{Stellar signal components seen in HARPS and HARPS-N\\solar radial velocities}

   %Author
   \author{K. Al Moulla\inst{1}
      \and X. Dumusque \inst{1}
      \and P. Figueira \inst{1,2}
      \and G. Lo Curto \inst{3}
      \and N. C. Santos\inst{2,4}
      \and F. Wildi    \inst{1}
    }

   %Institute
   \institute{Observatoire Astronomique de l'Université de Genève, Chemin Pegasi 51, 1290 Versoix, Switzerland\\
       \email{khaled.almoulla@unige.ch}
         \and Instituto de Astrofísica e Ciências do Espaço, Universidade do Porto, CAUP, Rua das Estrelas, 4150-762 Porto, Portugal
         \and European Southern Observatory, Avenida Alonso de Cordova 3107, Santiago de Chile, Chile
         \and Departamento de Física e Astronomia, Faculdade de Ciências, Universidade do Porto, Rua do Campo Alegre, 4169-007 Porto, Portugal
    }

   %Date
   \date{Received 2 August 2022 / Accepted 7 November 2022}

%Abstract
\abstract
% 5 {} token are mandatory, the first and last can be empty
% Context, optional
{\Acrfull*{RV} measurements induced by the presence of planets around late-type stars are contaminated by stellar signals that are of the order of a few meters per second in amplitude, even for the quietest stars. Those signals are induced by acoustic oscillations, convective granulation patterns, active regions co-rotating with the stellar surface, and magnetic activity cycles.}
% Aims
{This study investigates the properties of all coherent stellar signals seen on the Sun on timescales up to its sidereal rotational period. By combining HARPS and HARPS-N solar data spanning several years, we are able to clearly resolve signals on timescales from minutes to several months.}
% Methods
{We use a \acrfull*{MCMC} mixture model to determine the quality of the solar data based on the expected airmass-magnitude extinction law. We then fit the velocity power spectrum of the cleaned and heliocentric RVs with all known variability sources, to recreate the RV contribution of each component.}
% Results
{After rejecting variations caused by poor weather conditions, we are able to improve the average intra-day \acrfull*{RMS} value by a factor of ${\sim}1.8$. On sub-rotational timescales, we are able to fully recreate the observed RMS of the RV variations. In order to also include rotational components and their strong alias peaks introduced by nightly sampling gaps, the alias powers are accounted for by being redistributed to the central frequencies of the rotational harmonics.}
% Conclusions, optional
{In order to enable a better understanding and mitigation of stellar activity sources, their respective impact on the total RV must be well-measured and characterized. We are able to recreate RV components up to rotational timescales, which can be further used to analyse the impact of each individual source of stellar signals on the detectability of exoplanets orbiting very quiet solar-type stars and test the observational strategies of RV surveys.}

%Keywords

%Author, title
\authorrunning{K. Al Moulla et al.}
\titlerunning{Stellar signal components seen in HARPS and HARPS-N solar radial velocities}

%Titlepage
\maketitle

%MAIN TEXT

\section{Introduction}\label{Sect:1}

The external convective layer of late-type main-sequence stars give rise to several types of stellar signals seen in \acrfull*{RV} measurements, with timescales ranging from minutes to years \citep[see e.g.][]{Kjeldsen&Bedding1995, Saar&Donahue1997, Schrijver&Zwaan2000, Lindegren&Dravins2003, Christensen-Dalsgaard2004, Kjeldsen+2005, Meunier+2010, Dumusque+2011a, Dumusque+2011b, Meunier+2015, Dumusque2016}. The field of astroseismology, which studies stellar oscillations induced by acoustic waves propagating in the convective layer of late-type stars, has been established for a long time \citep{Goldreich&Keeley1977}, and is useful in the mass and age determination of main-sequence stars, and mass and radius determination of giant stars. More recently, the understanding of stellar signals has extended its importance to the field of exoplanets. Since the first discovery of an extra-solar planet orbiting a solar-like star by \cite{Mayor&Queloz1995}, more than $7800$ candidates have been detected, and the planetary status has been confirmed for $5200$ of them\footnote{\url{http://exoplanet.eu} (as of Oct. 10, 2022)}. A significant fraction of those planets have been either detected or confirmed using the RV technique, which allows to confirm the presence of planets by measuring the Doppler shift they induced on their host stars' spectra.

For state-of-the-art spectrographs which are able to reach sub-m/s RV precision, stellar activity is the main limitation to the detection of Earth-like planets, but also more massive planets at larger separation, as it induces RV variations that are one to two orders of magnitude larger than such planets. It is therefore crucial to better understand the different types of stellar signals, and finding mitigation techniques to correct RVs from them. Our understanding of the different types of stellar signals for solar-like stars was vastly improved with the HARPS spectrograph \citep[][]{Mayor+2003} at the ESO 3.6m Telescope, which was the first instrument to reach the m/s-precision level and therefore was able to measure all the different contributions of stellar signals, from the short-term acoustic oscillations, to the long-term magnetic cycles.

On the timescale of minutes, the excitation of acoustic oscillations, known as pressure waves (or p-modes), expands and contracts the stellar surface with different types of modes \citep[e.g.][]{Kjeldsen&Bedding1995, Christensen-Dalsgaard2004, Kjeldsen+2005}. These oscillations have been shown to be easily averaged out by selecting an observing strategy with a sufficiently long (about $5$ minutes for the Sun) exposure time \citep[e.g.][]{Dumusque+2011b, Chaplin+2019}.

On timescales from dozens of minutes to a few days, granulation phenomena on various spatial scales \citep[granulation, supergranulation, e.g.][]{DelMoro2004, DelMoro+2004}, driven by the convective motion in the external envelopes of solar-like stars, induce an RV signal with an amplitude of a few m/s \citep[][]{Dumusque+2011b, Cegla+2013, Meunier+2015}. Those signals can be averaged out in RV measurements by observing a star several times per night \citep[][]{Dumusque+2011b}, in order to sample different phases of the granulation phenomena signal, however, such a strategy cannot fully mitigate the supergranulation signal due to its longer than a day timescale, as demonstrated in \citet{Meunier+2015}. There is however hope that granulation phenomena signals can be mitigated by precisely measuring spectral line shape variations \citep[][]{Cegla+2019}.

RV modulations of a few m/s on the stellar rotational timescale originate from active regions co-rotating with the stellar surface \citep[e.g.][]{Saar&Donahue1997, Meunier+2010, Dumusque+2011a, Dumusque+2014, Herrero+2016}. These regions appear in conjunction with strong magnetic surface fields. They manifest themselves as either spots, which are regions of strong magnetic field and low temperature contrast compared to the quiet star and thus appear dark, or as faculae, which are regions of weaker magnetic field and of slightly higher temperature than the quiet star, usually surrounding spots but also found in isolation across the disk, as seen on disk-resolved images of the Sun. Spots and faculae introduce RV variations that can be decomposed to first order into two separated effects \citep{Meunier+2010, Bauer+18}. The first effect arises from the flux imbalance caused by active regions as they cross the oppositely Doppler shifted hemispheres of a star (unless it is seen pole on). The second effect arises from the suppression of convective blueshift by the strong magnetic fields of the active regions. Both of these effects cause quasi-periodic RV variations as the regions form at various latitudes (potentially depending on the phase of the magnetic cycle as in the case of the Sun) and the variations can sometimes have stronger peaks at half the stellar period if the regions rotate in and out of view from the visible hemisphere (for stars with an inclination close to $90^{\circ}$).

In Sect.~\ref{Sect:2}, we describe the solar telescopes and the RV data sets they have produced. In Sect.~\ref{Sect:3}, we assess the quality of the RV data through a mixture model based on the expected airmass-magnitude extinction law, and show how the intraday scatter is reduced by only considering points with high quality. In Sect.~\ref{Sect:4} we compute the \acrfull*{VPSD} and fit its identified components using a subset of the RV data covering sub-rotational timescales. We then recreate the RVs of the fitted components, and show that we fully recover the observed \acrfull*{RMS} variations, as a proof of concept of our methods. In Sect.~\ref{Sect:5}, we extend our methods to also include rotational timescales using the entire data sets, and in Sect.~\ref{Sect:6} we discuss and conclude on the obtained results and their implications.

\section{Solar observations with HARPS and HARPS-N}\label{Sect:2}

Due to the limitation induced by perturbing stellar signals when we want to detect small-mass exoplanets, it is essential to better understand the different types of stellar signals; in particular their respective sources of origin and their intrinsic properties. This will allow us to design optimal mitigation techniques to clean RV measurement from these perturbations. However, to improve our understanding of stellar signals using real observations is a challenging task, due to the limited and irregular sampling of night-time observations.

The difficulty in analysing stellar signals using stellar observations led to the HARPS-N solar telescope project: a small solar telescope, hosted at the 3.58m TNG Telescope at the Roque de los Muchachos Observatory in La Palma, Spain, that integrates the light from the solar disc and feeds it to the HARPS-N spectrograph, providing Sun-as-a-star RV measurements \citep[][]{Dumusque+2015, Phillips+2016}. Such a setup has many advantages. The Sun is a typical star among the targets that are followed-up to search for exoplanets, and since we are using the same spectrograph, anything that we learn from the Sun could in principle be extended to stellar observations (for G and likely K-dwarfs). Using HARPS-N during the day prevents competing with night-time observations, providing several hours of observation per day, on every clear day, which give us an exceptional sampling to probe stellar signals from minute to year timescales. Additionally, satellites like the \acrlong*{SDO} \citep[SDO,][]{Pesnell+2012} continuously observe the solar surface at high-spatial resolution in different wavelengths, but also in velocity space, and comparing what is happening in the RV measurements and simultaneously at the level of the solar surface can give us clues on the origin of the different types of stellar signals \citep[e.g.][]{Haywood+2016, AlMoulla+2022, Ervin+2022}.

HARPS-N is observing the Sun daily since July 2015. The HARPS-N solar telescope and the goal of the project is described in \citet{Dumusque+2015} and \citet{Phillips+2016}. Various important scientific results have already come out from the obtained data, on the side of data reduction \citep[e.g.][]{CollierCameron+2019, Dumusque+2021}, understanding of solar activity \citep[e.g.][]{Maldonado+2019, Milbourne+2019, Milbourne+2021} and data analysis to mitigate stellar activity in RV measurements \citep[e.g.][]{CollierCameron+2021, deBeurs+2022, Langellier+2021}.

The \acrfull*{HELIOS} is a copy of the HARPS-N solar telescope hosted at the ESO 3.6m Telescope at the La Silla Observatory in Chile, and feeds sunlight into HARPS since Sep. 8, 2018\footnote{\url{https://www.eso.org/public/announcements/ann18033/}}. HELIOS is a project led by the University of Geneva (PI: X. Dumusque) and the University of Porto (co-PI: P. Figueira). HARPS-N being a copy of HARPS, we expect very similar RV precision on the Sun between the HARPS-N and HARPS solar data sets. However, a significant difference comes from the fact that HARPS observes the Sun every minute, compared to every five minutes for HARPS-N, which allows us to resolve solar p-mode oscillations. With both instruments, the Sun is observed as long as possible each day, and as long as its altitude is higher than 10 degrees above the horizon.

Combining the solar observations taken with HELIOS and the HARPS-N solar telescope, the data analysed in this paper has the following properties. The HARPS data set consists of 376 observed days between Sep. 8, 2018 and Mar. 20, 2020 with a cadence of \SI{1}{min} (\SI{30}{s} exposure time and \SI{30}{s} read-out time). The HARPS-N data set consists of 745 observed days between Jul. 16, 2015 and Jul. 15, 2018 with a cadence of \SI{5.5}{min} (\SI{5}{min} exposure time and \SI{30}{s} read-out time).

\section{Data quality}\label{Sect:3}

\begin{figure*}[t!]
	\includegraphics[width=\textwidth]{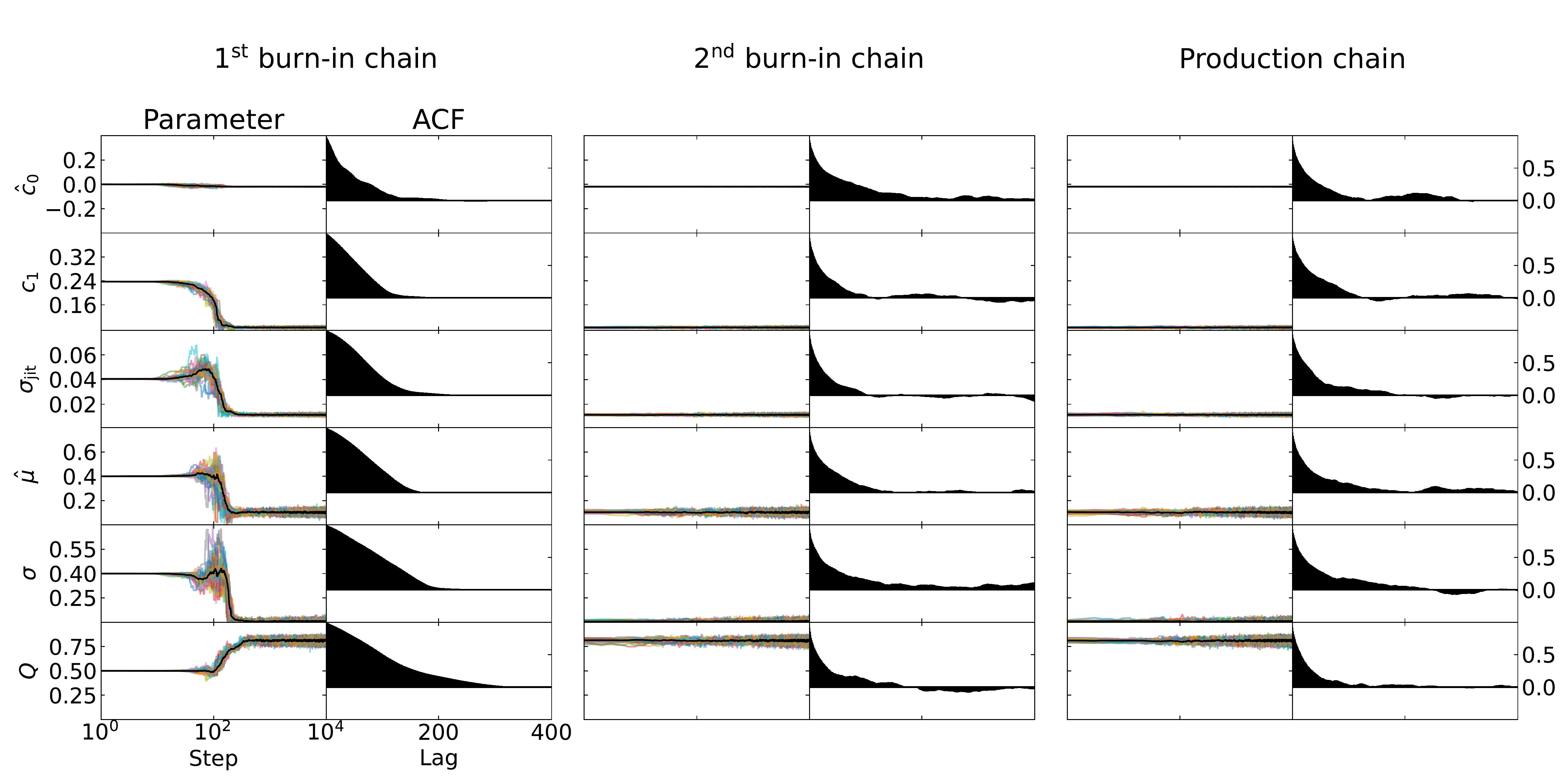}
	\caption{Example of MCMC parameter chains for HARPS observations taken on Nov. 11, 2019. The grouped columns represent the first burn-in chain (left), second burn-in chain (middle) and production chain (right), respectively. The left panels in each column show the sampled parameter values at each iterated step, where each coloured lines represent one walker and the the black line represents the median of all 32 walkers. The parameter axis limits are set equal to the bounds of their uniform prior distributions (see Table~\ref{Tab:1}). The right panels of each column shows the ACF up to a lag of 400 steps.}
	\label{Fig:1}
\end{figure*}

\begin{figure*}[t!]
	\includegraphics[width=\textwidth]{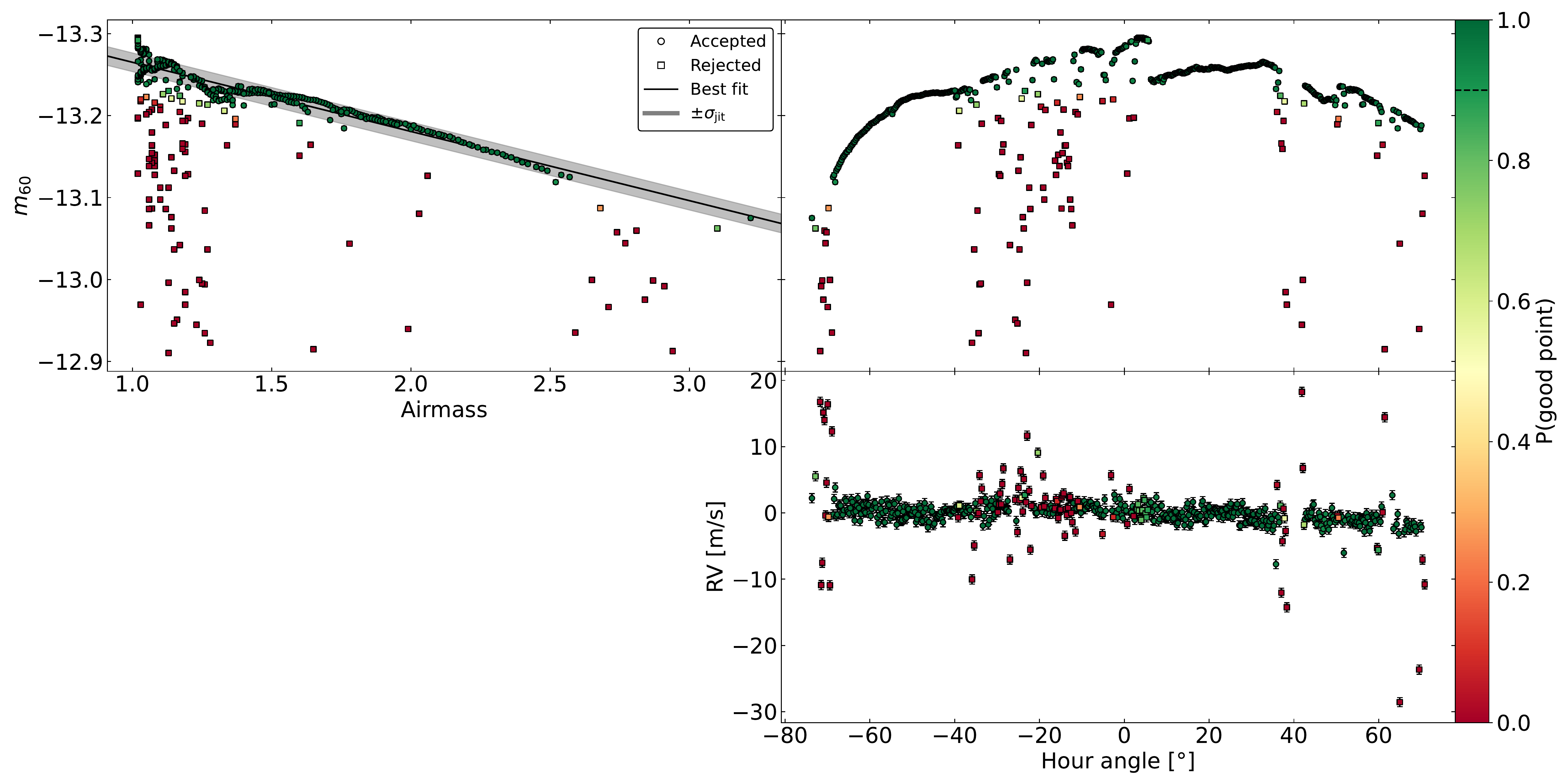}
	\caption{Example of quality assessment for HARPS observations taken on Nov. 11, 2019. \textit{Top panels}: $60$\textsuperscript{th} spectral order magnitude, $m_{60}$, as a function of airmass (left) and hour angle (right). The black line and grey area show the best fit and ${\pm}1\sigma$ interval for the estimated jitter. Data points are color-coded with their assessed quality shown in the colorbar. Points with a quality flag ${>}0.9$ are accepted (circles), and the rest are rejected (squares). The magnitude errorbars are smaller than the markers. \textit{Bottom panel}: RV as a function of hour angle. Note that the displayed RVs are still uncorrected for differential extinction.}
	\label{Fig:2}
\end{figure*}

\begin{figure*}[t!]
	\includegraphics[width=0.49\textwidth]{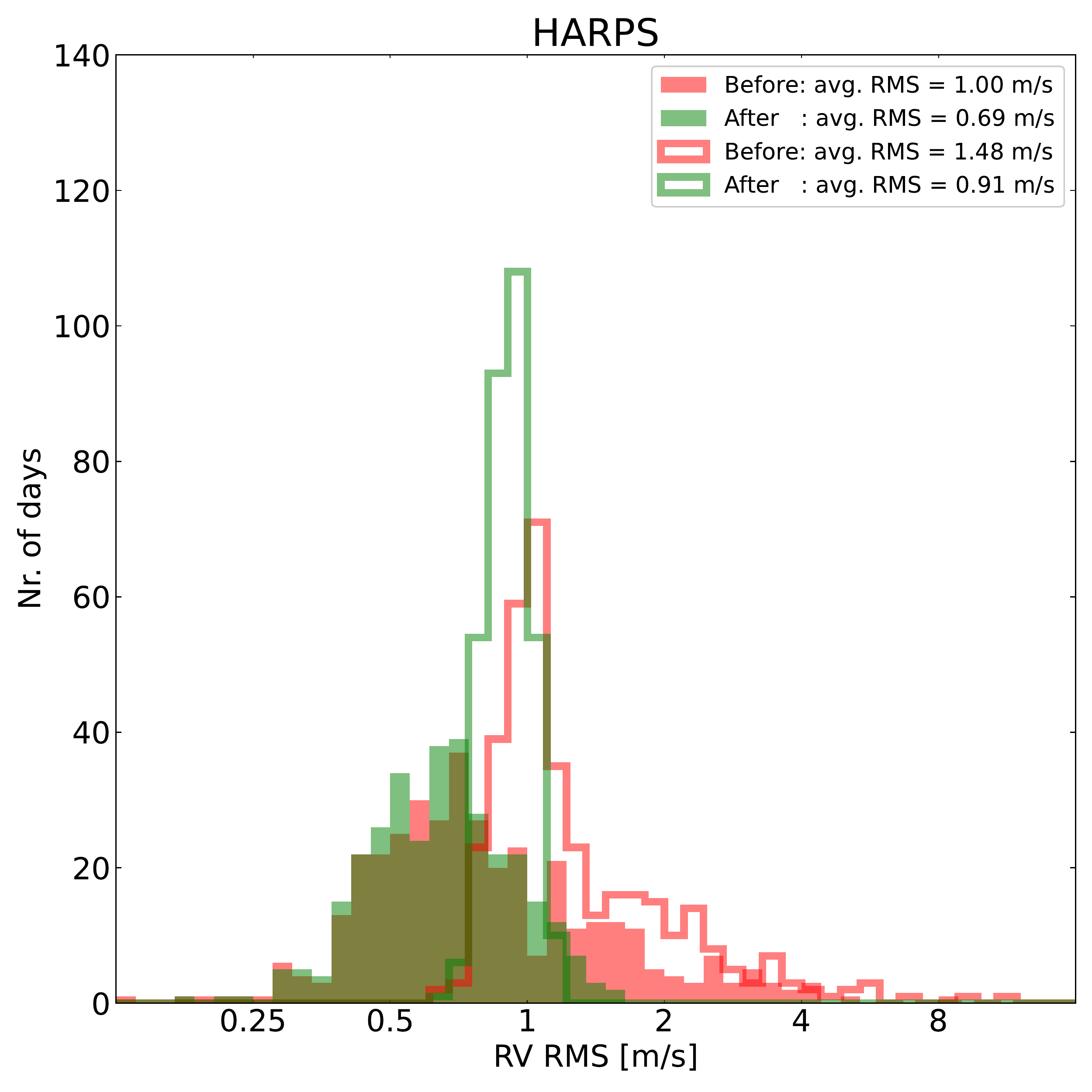}
	\includegraphics[width=0.49\textwidth]{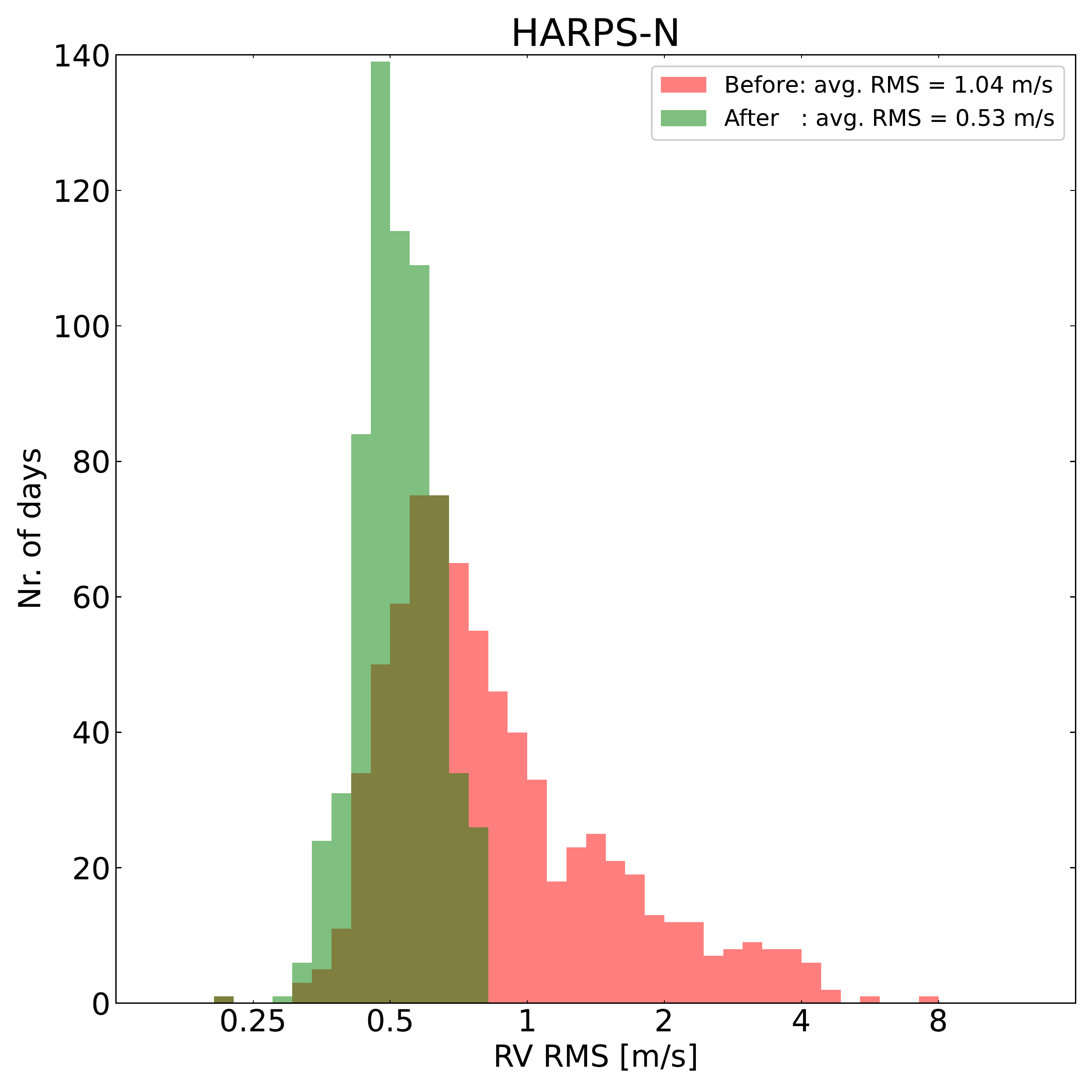}
	\caption{Improvement of daily RV RMS for HARPS (left panel) and HARPS-N (right panel). The histograms show the distribution of intraday RMS scatter before (in red) and after (in green) removing points with quality flags ${<}0.9$ and correcting for differential extinction with the fitted airmass-magnitude line parameters. For HARPS, we also binned the data every \SI{5.5}{min} to equal the sampling of HARPS-N data and with it average-out the signal induced by stellar oscillations; the RMS values of the binned HARPS data are shown in the fully colored histograms, and the RMS values of the unbinned HARPS data are shown in the step histograms. The legend displays the average RMS for the different distributions. Days with less than 10 measurements remaining after imposing the quality flag threshold are rejected. For HARPS, the number of included days decreased from 376 to 326; for HARPS-N, the number of included days decreased from 745 to 644.}
	\label{Fig:3}
\end{figure*}

Without simultaneous meteorological monitoring, the solar RV observations will occasionally be influenced by clouds and, in the case of HARPS-N located in La Palma, by dust winds from the Sahara called calima. These weather phenomena introduce one-sided outliers in the linear relation expected between the apparent magnitude of the Sun and its airmass as it traverses the sky \citep[Bouguer's Law;][]{Bouguer1729}, known as extinction. This has a chromatic effect, which will change the color balance of the observed spectrum, especially at high airmass where flux at shorter wavelengths gets absorbed, introducing an RV bias. Generally, for other stars, this is corrected at first order using an \acrfull*{ADC} and in a second step by normalizing the continuum of extracted spectra relative to simulated spectra reproducing the flux outside the atmosphere. For our solar observations, the ADC is not used and not needed since the chromatic images of the Sun fall on the integrating sphere, and since the color imbalance of the spectrum is mainly created by the fiber attenuation being chromatic and much stronger in the blue. However, due to the high \acrfull*{SNR} providing sufficient flux at all wavelengths, including the bluest part of the spectrum, we are able to balance the color even without an ADC. In addition to extinction, because the Sun is spatially resolved, it is also affected by differential extinction, caused when light rays from a resolved source traverse different airmass. Due to the tilt of the Sun’s rotation axis relative to the horizon normal, the differential extinction causes a flux inbalance between the red- and blueshifted hemispheres. To also correct for this, it is of importance to have an unbiased value of the extinction slope.

As a consequence of the non-gaussian nature of the noise, the method of least-square fitting becomes biased and insufficient. To accurately identify outlier points, we instead make use of a \acrlong*{MCMC} \citep[MCMC, e.g.][]{Ford2006, Hogg+2010} algorithm, which provides the ability to simultaneously model a fore- and background population with the use of a mixture model \citep{Foreman-Mackey2014}. In our case, the foreground population is represented by the linear airmass-magnitude relation, and the background population is represented by a gaussian noise model which is independent of the foreground. We follow the model framework of \cite{CollierCameron+2019}, by defining the following fore- and background likelihoods (the parameters $\theta_i$ are described in Table~\ref{Tab:1}), distinguished by the binary flag $q_i$ indicating whether a point belongs to the foreground ($1$) or background ($0$):
\begin{subequations}
\begin{align}
	p(y_i \,|\, x_i, \sigma_i, \boldsymbol{\theta}, q_i{=}1) = &\frac{1}{\sqrt{2\pi(\sigma_i^2+\theta_3^2)}} \nonumber \\
	&\exp \left(-\frac{(y_i - \theta_1 - \theta_2 x_i)^2}{2(\sigma_i^2+\theta_3^2)}\right) \,,           \label{Eq:1a} \\
	p(y_i \,|\, x_i, \sigma_i, \boldsymbol{\theta}, q_i{=}0) = &\frac{1}{\sqrt{2\pi(\sigma_i^2+\theta_5^2)}} \nonumber \\
	&\exp \left(-\frac{(y_i - \theta_4)^2}{2(\sigma_i^2+\theta_5^2)}\right)                \,,           \label{Eq:1b}
\end{align}
\end{subequations}
where $x_i$ are the airmass points, $y_i$ and $\sigma_i$ are the apparent magnitudes and their associated uncertainties given by the SNR in the 60\textsuperscript{th} spectrograph spectral order, chosen for its high SNR value. It is then possible to marginalize over the binary flags, in order to reduce the dimensionality of the problem, by introducing for them a simple prior,
\begin{equation}\label{Eq:2}
    p(q_i) =
    \begin{cases}
    Q   & \text{if} \, q_i = 1 \\
    1-Q & \text{if} \, q_i = 0
    \end{cases}
\end{equation}
where $Q$ can be seen as the fraction of good points for a given day. In this way, it is possible to find posterior constraints for the parameters $\theta$ and $Q$ by sampling the marginalized likelihood. More importantly, for each point we can find an estimate for the probability of being a fore- or background point, by marginalizing over the likelihood parameters,
\begin{equation}\label{Eq:3}
    p(q_i\,|\,y) \approx \frac{1}{N}\sum_{n=1}^{N} p(q_i\,|\,y,\,\theta^{(n)}) \,,
\end{equation}
where $n$ are the indices of the $N$ times sampled parameters.

\input{Tables/Tab_01.tex}

The numerical sampling is performed with the Python package \texttt{emcee} \citep{Foreman-Mackey+2013}. We adopt the number of iterations and sampling chains from \cite{CollierCameron+2019}, which are briefly described here. The chains consist of two burn-in chain, which are intended to facilitate the convergence of a third final production chain. Each chain has 32 so-called `walkers' per parameter which are iterated for 10,000 steps. In the first burn-in chain, the walkers of each parameter are initiated at the center of the uniform priors (see Table~\ref{Tab:1}) with an individual small perturbation. The walkers of the following two chains are initiated from the median of every 100\textsuperscript{th} step of the last 2000 steps of the preceding chain.

Figs.~\ref{Fig:1} and \ref{Fig:2} show the outcome of the MCMC analysis for one day (Nov. 11, 2019) observed with HARPS which had a significant amount of outliers. Fig.~\ref{Fig:1} shows the sampled parameter values of the walkers during the two burn-in and the production chains, along with the \acrfull*{ACF} with a lag up to 400 steps. We note that all the walkers converge within the first 1000 steps of the first burn-in chain, whereas the ACF in the following chains shows a decreased correlation between neighbouring steps. Fig.~\ref{Fig:2} indicates the outliers for this chosen day as seen in various variables. In the airmass-magnitude plot (from which the outliers are identified) they are seen as primarily one-sided deviants from the fitted linear relation. As a function of hour angle, the outliers have a higher magnitude (lower brightness) than the arch traced by the sun as it rises and sets, and these same points correspond to RV measurements with a larger scatter than the remaining points. In the upper right panel of Fig.~\ref{Fig:2}, there is a discontinuity in the magnitude at hour angle $0^{\circ}$ due to a meridian flip by the alt-azimuth mount of the solar tracker.

The derived line slope of the airmass-magnitude relation, known as the extinction coefficient, is used to correct the RVs for differential extinction following the methodology of \cite{CollierCameron+2019}. We thereafter impose a quality threshold, defined as the estimated probability being a foreground point (see Eq.~\ref{Eq:3}), of $0.9$ for all measurements. Fig.~\ref{Fig:3} shows the overall improvement on the intraday RMS. By correcting for differential extinction and rejecting low-quality measurements, we are able to improve the average daily RMS by a factor of $1.63$ and $1.96$ for HARPS and HARPS-N, respectively. The higher average daily RMS of HARPS compared to HARPS-N is attributed to its higher cadence (\SI{1}{min} instead of \SI{5.5}{min}), enabling HARPS to resolve the p-mode oscillations of the Sun which are averaged out in the HARPS-N measurements. We note that for this analysis, we also binned the HARPS measurements every \SI{5.5}{min}, to equal the HARPS-N sampling and with it mitigate the signal induced by stellar oscillations. After correction, we note that the binned HARPS data set reaches an average daily RMS value of $\SI{0.69}{\meter/\second}$, while the HARPS-N data shows $\SI{0.53}{\meter/\second}$. A significant difference that is likely due to the fact that HARPS-N data are reduced with the recent ESPRESSO data reduction software \citep[][]{Dumusque+2021}, which is not the case for HARPS products. We are currently adapting this procedure for HARPS data. In Appendix~\ref{Sect:A}, the distribution of derived MCMC parameters are shown in Fig.~\ref{Fig:A1}. Apart from $c_{0}$ and $\mu_{\mathrm{bg}}$, which are both the scaled with the apparent magnitude of the Sun, none of the parameters indicate any correlation. The standard deviation of the background population, $\sigma_{\mathrm{bg}}$, shows a bimodality since the sampling is unable to constrain the parameter on days without outliers on which its range is set by the bounds of the prior. Fig.~\ref{Fig:A2} shows the number of outlier points per day, where HARPS points within \SI{5.5}{min} from each other are only counted as one point to make the numbers comparable with HARPS-N. For most numbers of outliers per day, HARPS-N has a larger fraction of affected days than HARPS, which is indicative of the more humid and cloudy weather of La Palma compared to La Silla.

The Python code implemented to compute the quality flags is publicly available on GitHub\footnote{\url{https://github.com/almoulla/quality_flag}}, and can be easily adaptable to any solar spectroscopic observations which have simultaneous recordings of the airmass and SNR.

\newpage

\section{Stellar components on sub-rotational timescales}\label{Sect:4}

\begin{figure}[t!]
	\includegraphics[width=\linewidth]{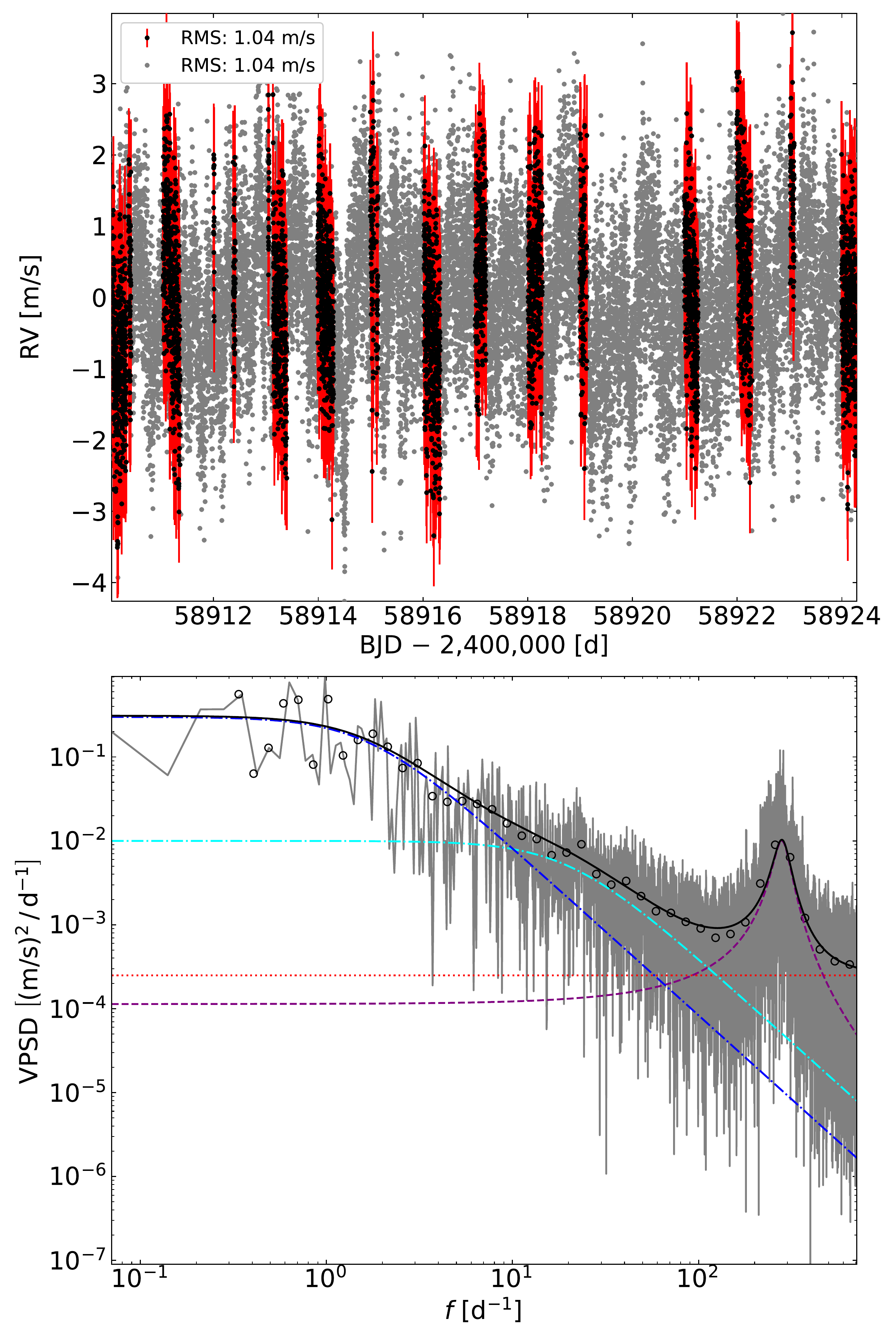}
	\caption{HARPS RVs and VPSD between Mar. 1 and 15, 2019. \textit{Top panel}: HARPS RVs for which the sub-rotational components are analyzed. The original observations are shown as black dots with red errorbars. The RVs recreated from the sum of all fitted VPSD components on an even time grid with the same cadence and end points as the observations are shown in grey. \textit{Bottom panel}: VPSD of the observed RVs (in grey) and the modeled components (colored curves), consisting of photon noise (dotted red), oscillations (dashed purple), granulation (dash-dotted cyan) and supergranulation (dash-dotted blue). The sum of models is optimized with respect to the black circles which correspond to the true VPSD averaged on logarithmically equidistant frequency steps.}
	\label{Fig:4}
\end{figure}

Almost all kinds of ground-based astronomical data suffer from uneven temporal sampling due to poor observing conditions and limited allocated telescope time affecting our ability to collect data, but most importantly due to the diurnal cycle imposed by Earth's rotation. And likewise, solar observations are subject to gaps due to nights. On long timescales (several months to years), these gaps introduce alias peaks in the Fourier spectra of temporal data \citep[e.g.][]{Dawson&Fabrycky2010}.

In Sect.~\ref{Sect:5}, the observed stellar components are analyzed on all timescales, up to and including the rotational one. However, as an intermediate step, we demonstrate here the validity of our method by first applying it on a sub-rotational timescale for which alias peaks and instrumental long-term drifts are less significant.

\subsection{Periodogram definition}\label{Sect:4.1}

We compute the \acrfull*{GLS} periodogram \citep{Zechmeister&Kurster+2009}, which fits the RVs with a model composed of a linear combination of sinusoidal functions including constant offsets:
\begin{equation}\label{Eq:4}
    \mathrm{RV}(t) = \sum_{\nu = 1/T}^{1/\left(2{\Delta}t\right)}a(\nu)\cos{2{\pi}{\nu}t} + b(\nu)\sin{2{\pi}{\nu}t} + c(\nu) \,,
\end{equation}
where $a(\nu)$, $b(\nu)$ and $c(\nu)$ are the fitted coefficients, and $\nu$ is the sampled frequencies ranging from $1/T$, where $T$ is the total time interval of the data, to the Nyquist frequency, $1/\left(2{\Delta}t\right)$, where ${\Delta}t$ is the median of the sampled time steps, in increments of $1/T$.

The GLS, as defined in \cite{Zechmeister&Kurster+2009}, returns for each frequency the normalised $\chi^2$ of the best-fitted model. To calculate the power of the signal at each frequency, what is known as the \acrfull*{VPS}, one calculates the summed squares of the sine and cosine terms:
\begin{equation}\label{Eq:5}
    \mathrm{VPS}(\nu) = a(\nu)^2 + b(\nu)^2 \,.
\end{equation}
In order to obtain the \acrfull*{VPSD}, which is independent of the observing window, we multiply the power by the effective length of the observing run, calculated as the inverse of the area under the window function \citep[in units of power, see][]{Kjeldsen+2005}. We note that our window function is obtained by computing the periodogram of a pure sine wave at the center-most sampled frequency. In Fig.~\ref{Fig:4}, we observe that the power decays from lower to higher frequencies, in a behaviour that is commonly attributed to granulation phenomena at several timescales; as the frequency increases, one identifies the emergence of the p-mode bump induced by multiple individual oscillation modes.

\subsection{Analytical functions for power spectrum components}\label{Sect:4.2}

In order to recreate the RV contribution of each stellar component, we need to separate their power spectrum counterparts. This can be done by fitting the VPSD using analytical functions that capture the contribution of each type of stellar signal. Using analytical functions also allows us to reconstruct the RV of each component on an evenly-sampled time grid. More generally, being able to reconstruct the RV on an evenly-sampled time grid is crucial to test different observational strategies to mitigate stellar signal \citep[e.g.][]{Dumusque+2011b, Dumusque2016}, or to estimate stellar contributions when searching for exoplanets \citep[e.g.][]{Meunier+2010, Dumusque+2017, Meunier+2019}.

We follow the methodology of \cite{Lefebvre+2008}, and chose to fit the different stellar signal components with two types of analytical functions. The periodic components are fitted with a Lorentz function of the form
\begin{equation}\label{Eq:6}
    \mathrm{VPSD}_{\mathrm{L}}(\nu) = A_{\mathrm{L}}\frac{\Gamma^2}{(\nu - \nu_0)^2 + \Gamma^2},
\end{equation}
where $A_{\mathrm{L}}$ is the amplitude, $\Gamma$ is the \acrfull*{FWHM}, and $\nu_0$ is the central frequency. The granulation phenomena are fitted with Harvey functions of the form
\begin{equation}\label{Eq:7}
    \mathrm{VPSD}_{\mathrm{H}}(\nu) = \frac{A_{\mathrm{H}}}{1 + (\tau\nu)^\alpha}
\end{equation}
where $A_{\mathrm{H}}$ is the amplitude, $\tau$ is the characteristic timescale, and $\alpha$ is the power-law slope.

The fitting procedure is done with the Python module \texttt{lmfit} \citep{Newville+2020}, which implements a Levenberg–Marquardt minimization. To reduce degeneracies from the large number of fitted parameters, we constrain our Harvey functions by setting $\alpha{=}2$ as in \cite{Lefebvre+2008}.

\subsection{Recreated sub-rotational RV components}\label{Sect:4.3}

To demonstrate the validity of our method on sub-rotational timescales, we chose to recreate the contributing RV components as seen by HARPS for a duration of 15 days (i.e., roughly half the solar sidereal rotation period). We select the part of the solar data set between Mar. 1 and 15, 2019 (see Fig.~\ref{Fig:4}). After computing the unnormalized VPSD, we bin the spectrum in 50 logarithmically equidistant frequency points and exclude the points at $\log_{10}{f}{<}-0.5$ where the sampling is too sparse. We then fit four components simultaneously: a constant photon noise level, one Lorentzian function for the envelope of oscillations, and two Harvey functions for granulation and supergranulation. The optimization is performed by minimizing the residual between the sum of all components and the binned true VPSD.

The individual VPSD components can be reconstructed into RVs using the following formula,
\begin{equation}\label{Eq:8}
    \text{RV}_{i}(t) = \sum_{\nu} \sqrt{\text{VPSD}_{i}(\nu)\Delta\nu}\sin(2\pi\nu t + \phi(\nu)) \,,
\end{equation}
where $t$ are the recreated times in step of ${\Delta}t$, $\nu$ and ${\Delta}\nu$ are the frequencies and frequency step of the VPSD, and $\phi(\nu)$ is a randomly chosen phase between $0$ and $2\pi$. The upper panel of Fig.~\ref{Fig:4} shows the real RV data selected for the short timescale window as well as the recreated RVs of the sum of the fitted components in the lower panel of the same figure. As indicated in the legend, both the real and recreated RVs have the same RMS, indicating that the fitted components are able to capture all significant stellar signals present on sub-rotational timescales. We are also able to recreate the individual RV contributions, as shown in Appendix~\ref{Sect:B}, where Fig.~\ref{Fig:B1} shows the RV amplitudes and RMS values of all fitted components compared to the true total signal.

\section{Stellar components up to rotational timescales}\label{Sect:5}

\subsection{Combining the HARPS and HARPS-N data sets}\label{Sect:5.1}

\begin{figure*}[t!]
	\includegraphics[width=\textwidth]{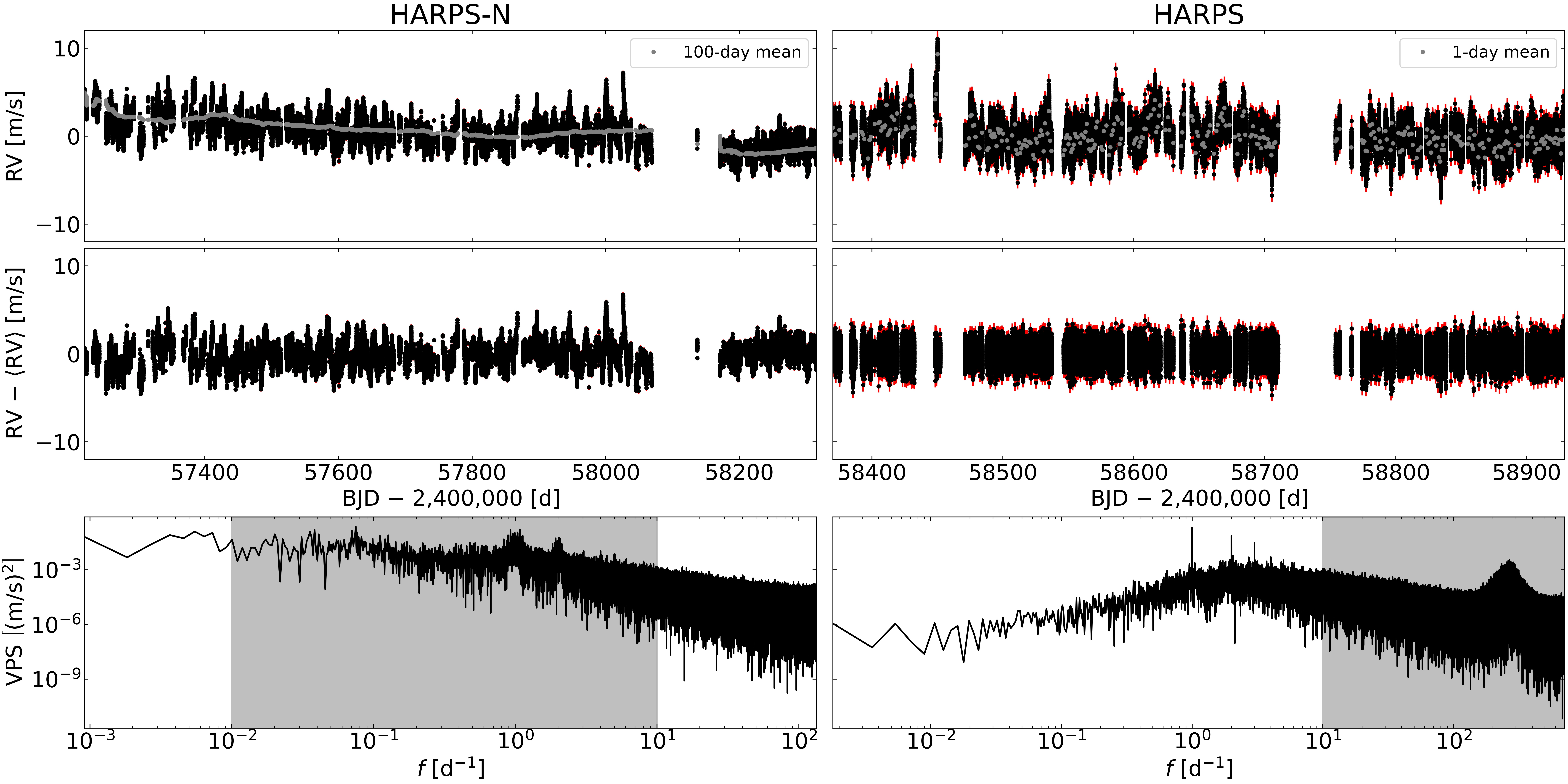}
	\caption{Total RV time series and VPS for HARPS and HARPS-N. \textit{Top}: RV time series. \textit{Middle}: HARPS-N RVs detrended by a 100-day rolling mean, and HARPS RVs detrended by a weighted daily mean. \textit{Bottom}: GLS periodogram where the gray regions are merged together.}
	\label{Fig:5}
\end{figure*}

\begin{figure*}[t!]
	\includegraphics[width=\textwidth]{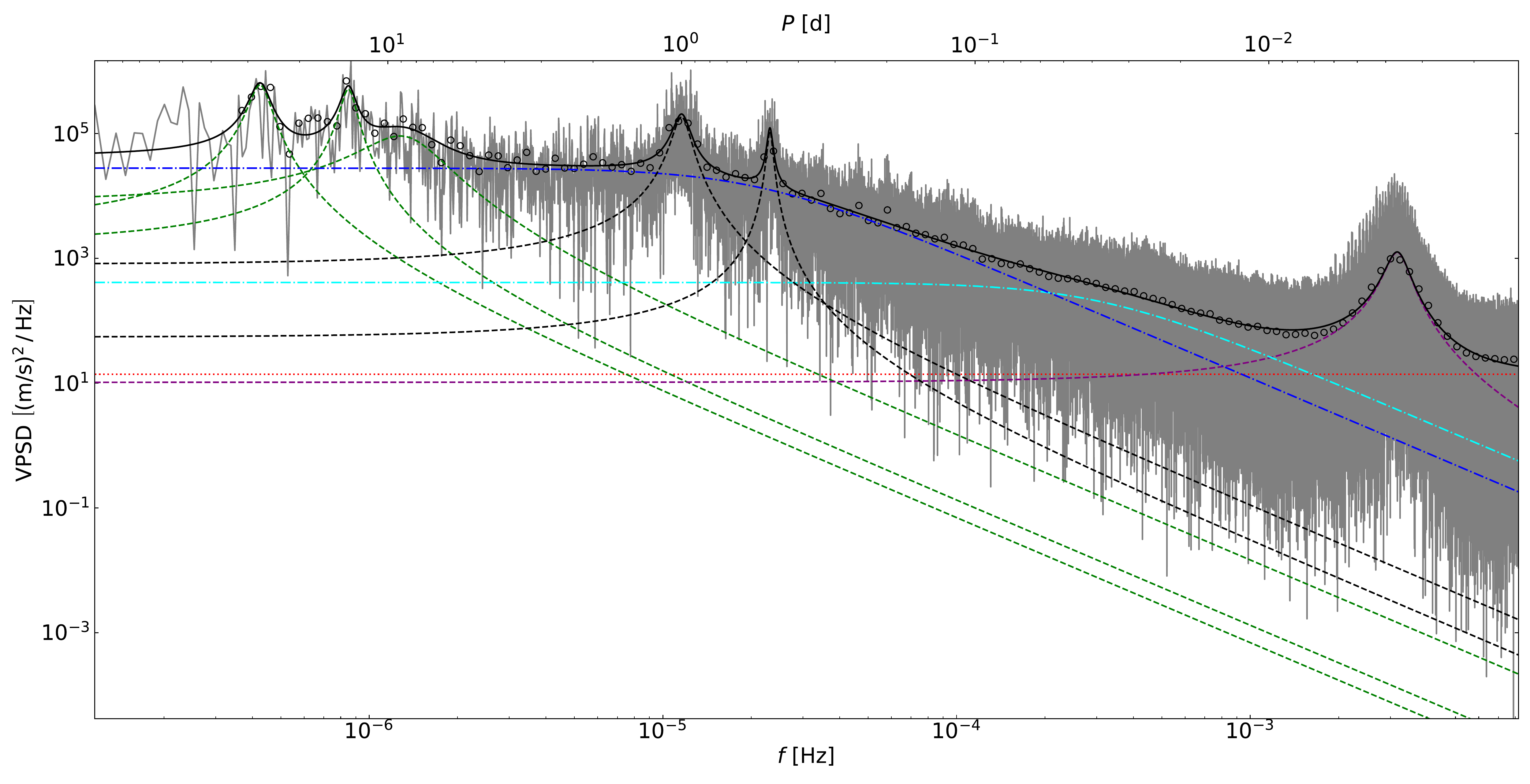}
	\caption{Stellar signal components fitted to the combined HARPS and HARPS-N VPSD (solid gray curve). The dotted red curve shows the photon-noise constant. The dashed curves represent Lorentzian functions for oscillations (purple), rotation and its harmonics (green), and aliases (black), whereas dash-dotted curves represent Harvey functions for granulation (cyan) and supergranulation (blue). The solid black curve is the sum of all components which was fitted with respect to the binned VPSD (black circles).}
	\label{Fig:6}
\end{figure*}

\input{Tables/Tab_02.tex}

The combination of HARPS measurements with minute-timescale cadence and HARPS-N measurements with long time coverage (3 years), enables us to resolve both low and high frequency stellar signals. By constructing a merged periodogram of the two data sets, we are able to simultaneously fit all signal components. We choose to merge the periodograms in the frequency domain, rather than merging the RV data sets and performing a periodogram on the joint set. The reasoning behind this is that unless the RV sets are fully overlapping, their different cadences will cause some stellar signals to be inconsistent in the time domain and furthermore our cut-off for the sampled frequencies will be a poor estimate of the true Nyquist frequency.

The individual RV data sets are detrended on non-interfering timescales adjusted for the different frequency domains each set is able to resolve. The HARPS measurements, which are used to resolve p-mode oscillations and small-scale granulation, is detrended with the inverse variance weighted daily mean in order to mitigate any day-to-day offsets induced by instability in daily wavelength solution derivation \citep[e.g.][]{Dumusque2018, Trifonov+2020, Dumusque+2021}. The HARPS-N measurements, which are used to resolve all other signals, i.e. large-scale granulation and rationally modulated activity, is detrended with a 100-day rolling mean in order to mitigate the 11-year magnetic activity cycle and instrumental long-term drifts. Fig.~\ref{Fig:5} shows the entire RV time series of both instruments before and after detrending.

The VPS and VPSD are then computed in the same way as in Sect.~\ref{Sect:4} on the detrended RVs. The VPSDs are then combined together by merging the HARPS-N VPSD in the frequency range $f{=}10^{-2}-10^{1}\,\mathrm{d}^{-1}$ with the HARPS VPSD in the frequency range $f{=}10^{-1}-f_{\mathrm{max}}\,\mathrm{d}^{-1}$, where $f_{\mathrm{max}}$ is the highest resolved frequency with HARPS. The lower limit is set by the 100-day detrending imposed on the HARPS-N RVs, and the frequency at which the VPSDs are merged is taken to be a rough mid-point between the optimal frequency windows of each detrended data set. We fit the same signal components to the VPSD as before, i.e. photon noise, oscillations, granulation and supergranulation, in addition to the rotational components.

\subsection{Recreated rotational RV components}\label{Sect:5.2}

The sub-rotational components are recreated as described in Sect.~\ref{Sect:4}, but on a different timescale. Since the frequency scale is bounded by the total time span of the HARPS-N observations and the cadence of the HARPS observations, we chose to now recreate the components on the timescale and cadence of only the HARPS observations. This is because the longer the timescale selected, the greater the number of computed time points and the number of frequencies over which the summation is performed in Eq.~\ref{Eq:8}, significantly increasing the non-linear computational time. The time span of the HARPS observations is nearly two years and thus large enough to cover many rotational periods.

The rotational contributions are quasi-periodic with a periodicity of the sidereal rotation period of the Sun of about ~27 days and its harmonics. We choose to model the fundamental and first two harmonics, and fit each component with a Lorentz function, alike the p-mode oscillations. To reduce the number of free parameters, we imposed that the frequencies of the first and second harmonics were twice and thrice the value of the fundamental.

The rotational harmonics with central frequencies $\nu_{0,i}$ ($i{=}0,1,2$) convolved with the window functions of HARPS-N, which has strong daily alias peaks due to solar observations being restricted to daytime, will create aliases in the VPSD at frequencies
\begin{equation}\label{Eq:9}
    \nu = |\nu_{0,i} \pm m \, \text{d}^{-1}| \,,
\end{equation}
where $m$ is an integer number. Since the central frequencies of the rotational peaks are much smaller than unity, the aliases appear at roughly $\nu {\approx} 1,2,\dots \, \text{d}^{-1}$. The forest of peaks around the two primary aliases are also fitted with Lorentzian functions with fixed centers. Finally, since some of the spectral power has been allocated from the rotational peaks to the aliases, we attempt to redistribute the alias power when recreating the rotational RV contribution. This is executed by first summing over the alias Lorentizians evaluated at the frequencies corresponding to the recreated time points. The summed alias VPSD is then proportionally divided between the rotational peaks, where each peaks receives a fraction
\begin{equation}\label{Eq:10}
    f_i = \frac{A_i}{\sum_i A_i} \,.
\end{equation}
where the index $i$ corresponds the $i^{\text{th}}$ rotational harmonic with amplitude $A_i$. The redistributed power is added at $\Tilde{\nu}_{0,i}$, the sampled frequency closest to the central frequency of the $i^{\text{th}}$ rotational component. The customization of Eq.~\ref{Eq:8} for the rotational contribution thus becomes
\begin{equation}\label{Eq:11}
\begin{split}
    \mathrm{RV}_{\mathrm{rotation}}(t) = &\sum_{\nu} \sqrt{\mathrm{VPSD}_{\mathrm{rotation}}(\nu)\Delta\nu}\sin(2\pi\nu{t} + \phi(\nu)) \, +\\
    &\sum_{i}\sum_{\nu} f_i\sqrt{\mathrm{VPSD}_{\mathrm{alias}}(\nu)\Delta\nu}\sin(2\pi\Tilde{\nu}_{0,i}{t} + \phi(\Tilde{\nu}_{0,i})) \,.
\end{split}
\end{equation}
The alias components are not reconstructed in the same way as the other components since, although they might contribute to the total RMS in agreement with the observed scatter, their reconstructed RVs will have too short periodicities since they will be related to the frequency of the sampling gaps instead of the stellar signal from which they originate.

\begin{figure*}[t!]
	\includegraphics[width=\textwidth]{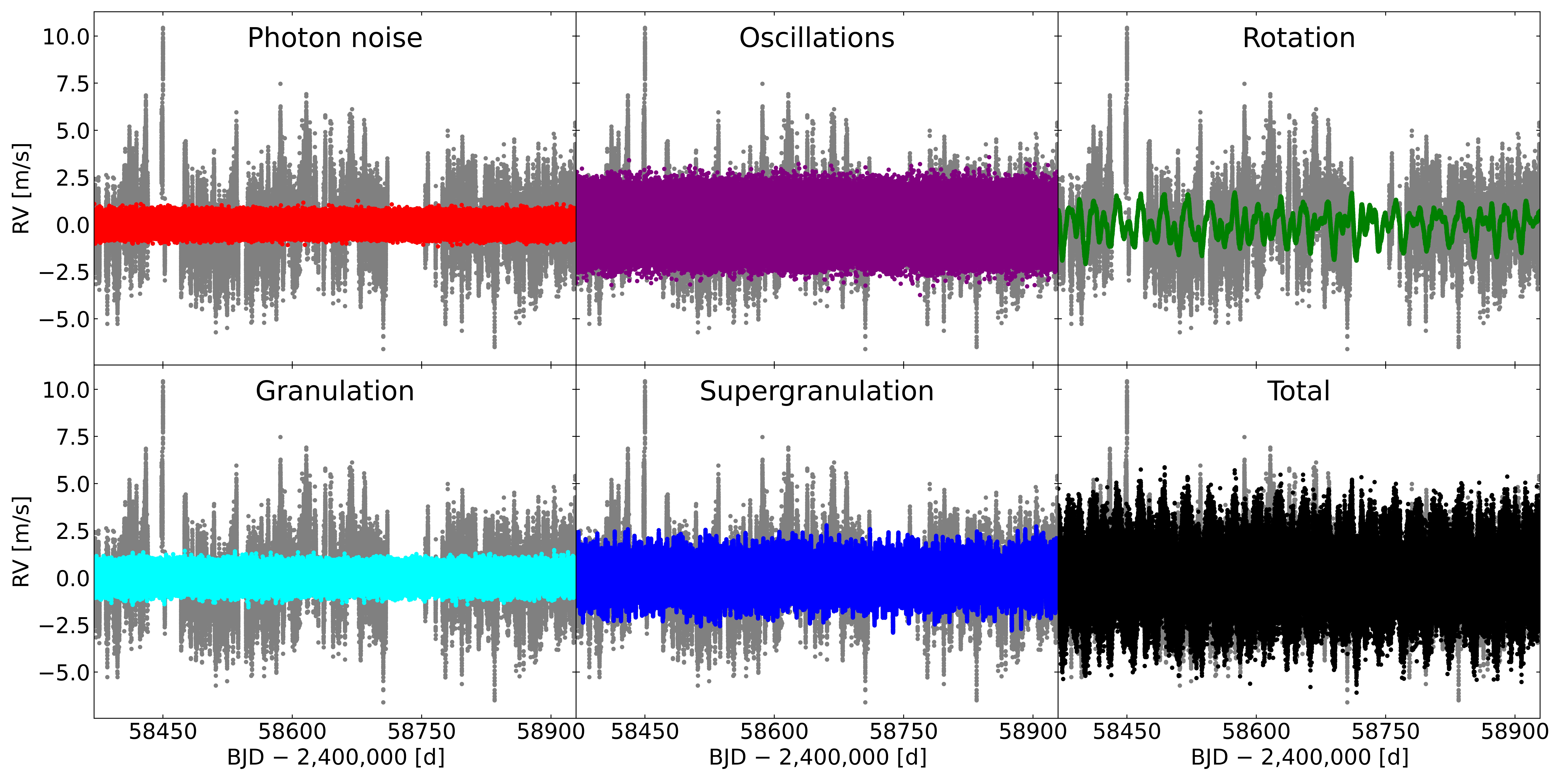}
	\caption{Recreated RV components up to rotational timescales. Individual RV contributions from the stellar signals fitted in the VPSD of Fig.~\ref{Fig:6}. The components are indicated in the top of each subplot and the RV points are color-coded according to the curve colors in Fig.~\ref{Fig:6}. The true total signal from the HARPS RVs (without errorbars) with a linear drift subtracted is shown in gray for comparison.}
	\label{Fig:7}
\end{figure*}

The fitted VPSD components are shown in Fig.~\ref{Fig:6}, and the best-fit values of the coefficients of the analytical expressions are listed in Table~\ref{Tab:2} together with their estimated uncertainties. The recreated individual RV components are shown in Fig.~\ref{Fig:7}, allowing the peak-to-peak comparison of various signals. The bottom-right panel shows the recreated RVs obtained from the sum of all components, which is able to capture similar structure and scatter as the true total signal. Table~\ref{Tab:3} displays the RV RMS of both the observed data sets and the recreated components. We note that on the original HARPS data, we also fitted for a linear, long-term trend to remove any signal induced by the solar magnetic cycle. The RV RMS obtained on HARPS measurements, $\SI{1.58}{m/s}$, is comparable to the one obtained with HARPS-N once the long-term trend is removed ($\SI{1.37}{m/s}$) and the contribution of the unresolved oscillations are added ($\sqrt{1.37^2+0.73^2}=\SI{1.55}{m/s}$).

Although the total recreated signal is visually similar in amplitude and modulation, its RMS value is about $\SI{27}{cm/s}$ lower than that of the HARPS data, on whose timescale the RVs were simulated. We believe this discrepancy could be due to several factors associated with attempting to extend this kind of analysis to rotational timescales. The most uncertain among them is the validity of our method when used to combine correlated contributions, such as the rotational harmonics. On sub-rotational timescales, the validity of using random phases when recreating p-mode oscillations and granulation phenomena was motivated by their stochastic and thus independent nature. However, by attributing the rotational harmonics random phases as well could lead to deconstructive interference at times when they are behaving more coherently in the real observations. Another possibly impacting factor is the influence of the solar magnetic cycle, which does not only affect the RVs with a long-term drift (which we corrected for by detrending) but also affect the intrinsic amplitudes and characteristic timescales of the modeled stellar signal with time.

\input{Tables/Tab_03.tex}

\section{Discussion and conclusions}\label{Sect:6}

An important step towards better understanding the origin and mechanisms of solar and stellar variability, is to accurately measure the magnitude and timescales of various activity sources, allowing subsequent models to build on better constrains from observations. In this study we have assessed the quality of high-precision, high-cadence solar RVs taken with the HARPS and HARPS-N spectrographs to build a near-continuous data set suitable for investigating the typical stellar signals which are present at all timescales up to the stellar rotation period. We have analysed the merged VPSDs of both instruments to highlight the best-resolved features from each one.

We have been able to show that using fairly simple analytical functions, one can easily estimate RV budget of various stellar signals at different frequency ranges. On sub-rotational timescales, we were able to fully capture the significant contributions of p-mode oscillations and granulation phenomena. We find that solar oscillations induce a RV RMS of $\SI{0.73}{m/s}$, which is fully compatible with the solar estimation performed in \citet[][see Fig.~3]{Chaplin+2019}. Regarding granulation and supergranulation, we find RV RMS contributions at the level of 0.31 and $\SI{0.68}{m/s}$, respectively. For granulation, the derived value is significantly smaller that the outcome of the simulation performed in \citet{Meunier+2015} which found $\SI{0.8}{m/s}$. The same study also simulate supergranulation and estimate a contribution to a RV RMS level between $0.28$ and $\SI{1.12}{m/s}$. The value that we found analysing solar data is at the middle of this range and therefore fully compatible. We note that the discrepancy found for granulation between observations and simulations could be due to different range in frequency considered as granulation induces signal over a large range of frequencies.

We were also able to extend our analysis up to rotational timescales through the inclusion of quasi-periodic peaks originating from active magnetic regions co-rotating with the solar surface. By modeling both the most prominent harmonics and their associated alias peak due to the uneven time sampling, and proportionally redistributing the integrated alias power to the central frequencies of the rotational components, we were able to reach similar peak-to-peak variations and slightly underestimated RV RMS when comparing the true signals and the sum of our fitted components. In our analysis, we find that rotational modulation induce a RV RMS of $\SI{0.74}{m/s}$ on the Sun during the minimum of its activity cycle.

The contribution of oscillations, supergranulation and rotational activity are at very similar RV RMS levels. However, they do not contribute equally when searching for exoplanets. Indeed, due to the short timescale of oscillations, their impact can be reduced to levels lower than $\SI{0.1}{m/s}$ by imposing exposure times longer than their typical timescale \citep[e.g.][]{Chaplin+2019}. This is not the case for supergranulation that evolves with a timescale of up to several days. Although taking several measurements per night can reduce the impact of this stellar signal \citep{Dumusque+2011b}, \citet{Meunier+2015} show that even after averaging data over 5 days, the RV RMS is above $\SI{0.17}{m/s}$. Regarding rotational activity, this stellar signal is even more complex to handle as we can no longer average out data to mitigate its impact.

Since our modeling of stellar signals can be used to generate RVs with a continuous and extremely high cadence, any realistic sampling can be reproduced and therefore the result of this paper can be used to test observational strategies, how they can mitigate stellar signals and influence the detection of exoplanets. It would be possible, for example, to optimise observations of further RV surveys aiming at detecting Earth-analogues orbiting very quiet stars.

As solar telescopes equipped with high-resolution spectrograph continue to gather near-continuous data, similar studies can be extended to cover one or several magnetic cycles. Future work could also investigate the temporal evolution of recreated RV components by splitting the available data according to the minima or maxima of such cycles.

%Acknowledgments
\text{}\\
\footnotesize
\noindent
\textit{Acknowledgments.} We thank the anonymous referee for valuable comments which helped improve the quality of the manuscript. This work has been carried out within the framework of the National Centre of Competence in Research PlanetS supported by the Swiss National Science Foundation under grants 51NF40\_182901 and 51NF40\_205606. The authors acknowledge the financial support of the SNSF. This project has received funding from the European Research Council (ERC) under the European Union’s Horizon 2020 research and innovation program (grant agreement SCORE No. 851555). NCS acknowledges the support from the European Research Council through the grant agreement 101052347 (FIERCE). This work was supported by FCT - Fundação para a Ciência e a Tecnologia through national funds and by FEDER through COMPETE2020 - Programa Operacional Competitividade e Internacionalização by these grants: UIDB/04434/2020; UIDP/04434/2020.

%References:
\balance
\bibliographystyle{aa}
\bibliography{References}

%Appendix
\onecolumn
\begin{appendix}

%Appendix

\section{Statistics of MCMC parameters and outliers}\label{Sect:A}

\begin{figure*}[h!]
    \includegraphics[width=\textwidth]{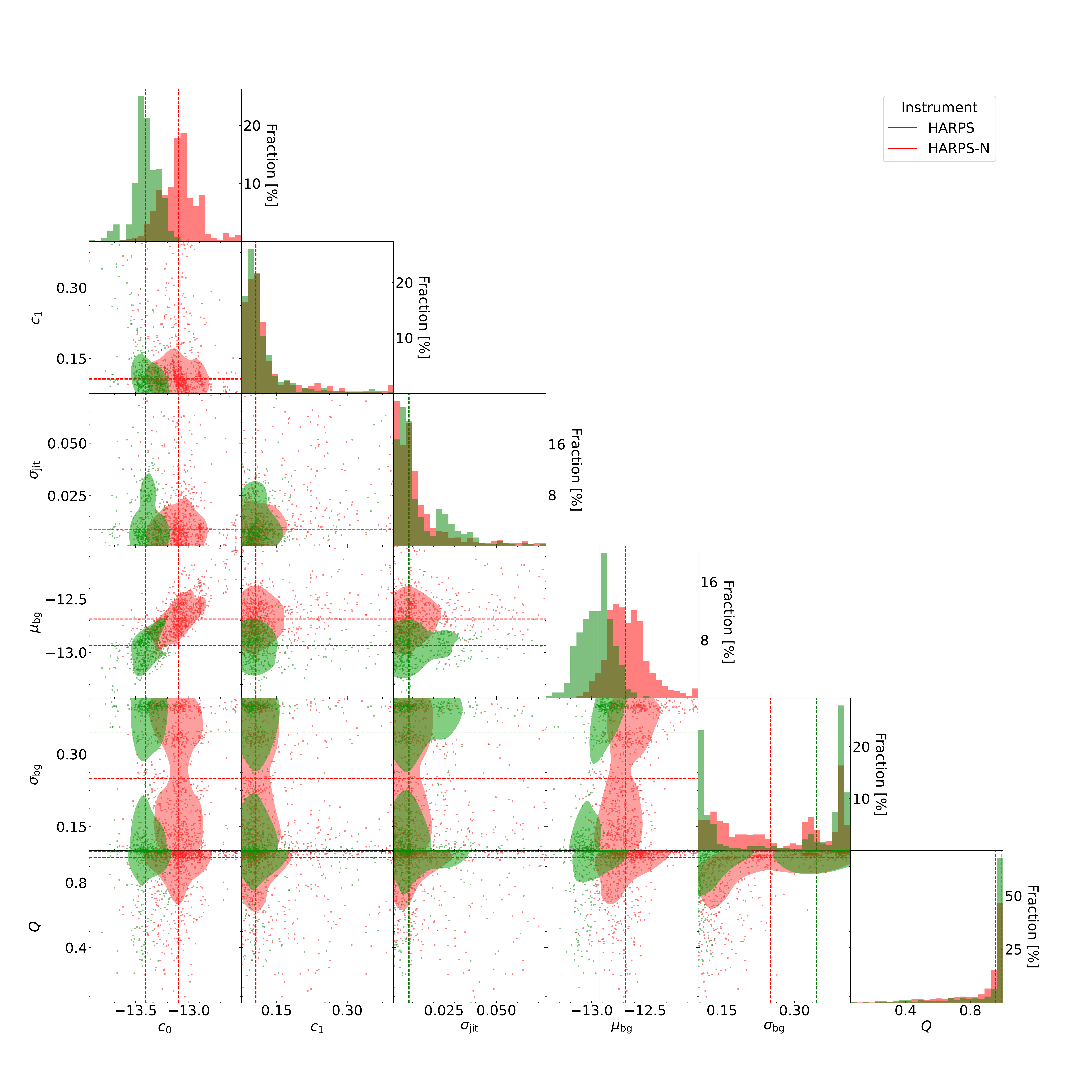}
    \caption{Corner plot of MCMC parameters. The subplots show the distribution of the daily fitted parameters described in Table~\ref{Tab:1}, for both HARPS (green points) and HARPS-N (red points). The top subplot of each column shows the histogram in units of fraction of the total number of days per instrument. The dashed lines indicate the median values, and the contour curves represent the ${\pm}1\sigma$ kernel densities.}
     \label{Fig:A1}
\end{figure*}

\begin{figure*}[h!]
    \centering
    \includegraphics[width=0.5\textwidth]{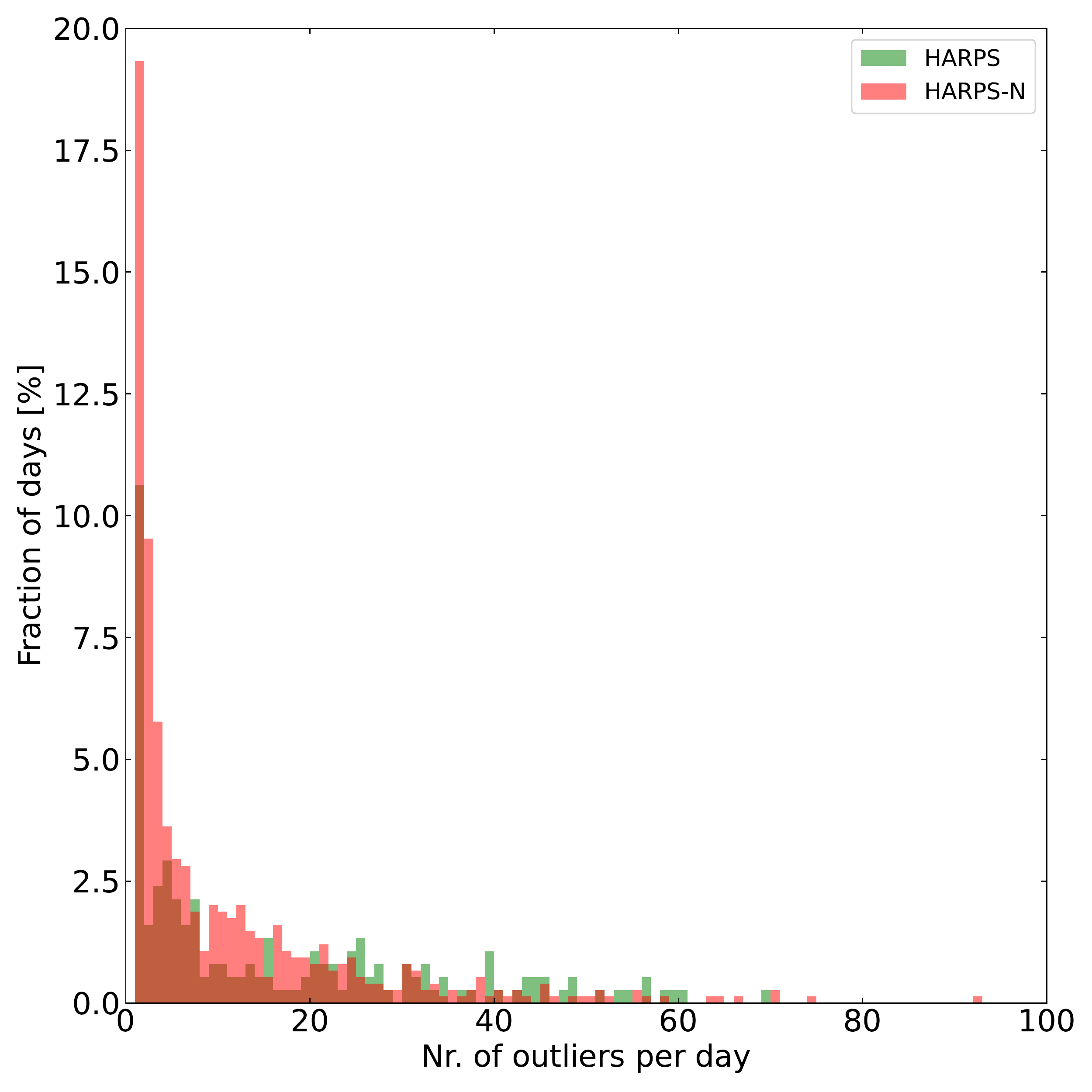}
    \caption{Outlier statistics. The histograms show the distribution of the number of outliers, i.e. points with a quality flag ${<0.9}$, per day for HARPS (green) and HARPS-N (red). The histograms are binned between 1 and 100 outliers per day in steps of 1. The fraction of days is relative to the number of days observed with each instrument. For HARPS and HARPS-N, $47.3\%$ and $76.1\%$, respectively, of all observed days have at least one outlier.}
     \label{Fig:A2}
\end{figure*}

\newpage

\section{RV components on sub-rotational timescales}\label{Sect:B}

\begin{figure*}[h!]
    \includegraphics[width=\textwidth]{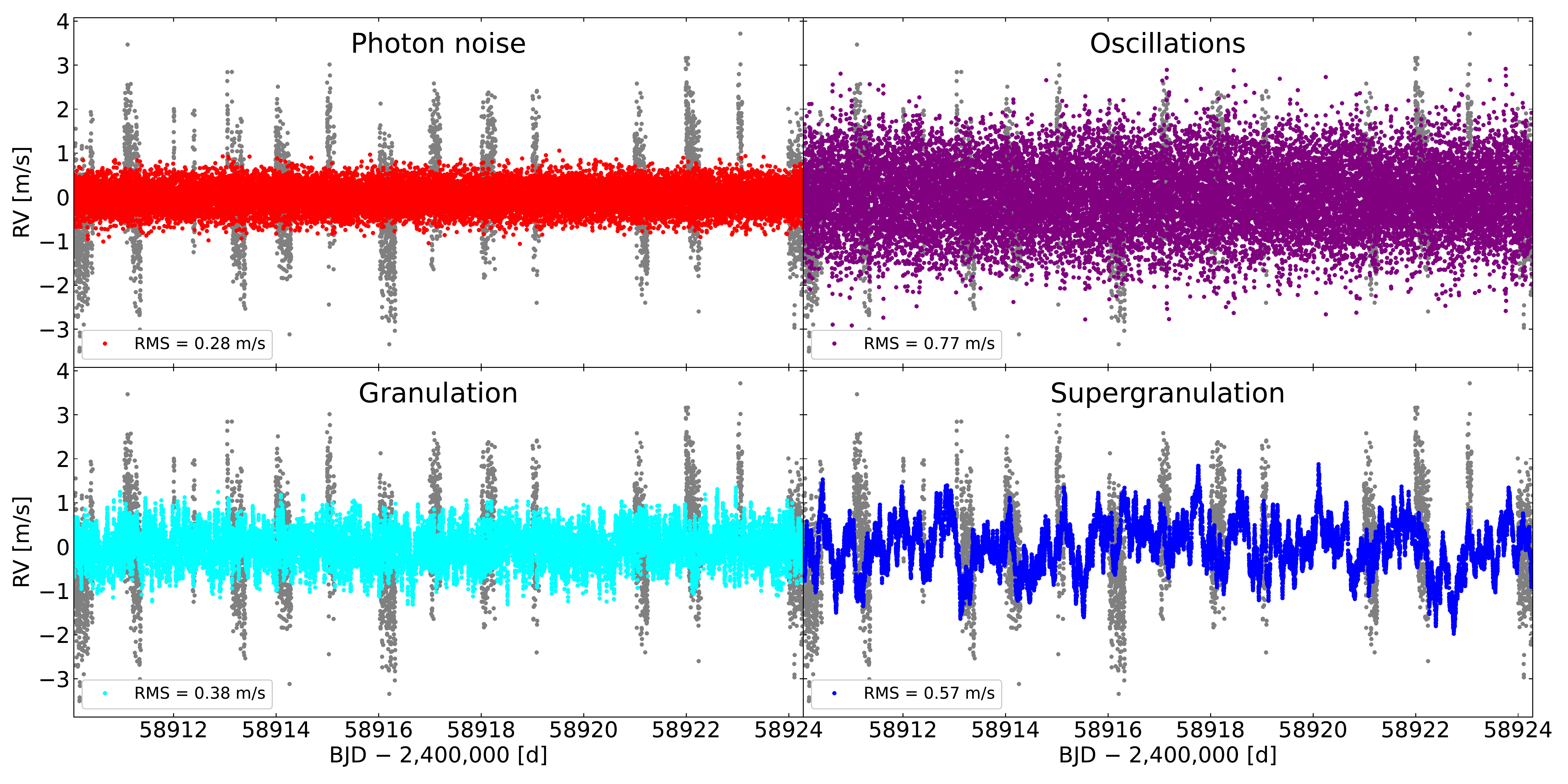}
    \caption{Recreated RV components on sub-rotational timescales. Individual RV contributions from the stellar signals fitted in the VPSD of Fig.~\ref{Fig:4}. The components are indicated in the top of each subplot and the RV points are color-coded according to the curve colors in Fig.~\ref{Fig:4}. The RMS of each component is shown in the lower left legend in each subplot. The true total signal (without errorbars) is shown in gray for comparison.}
    \label{Fig:B1}
\end{figure*}

\end{appendix}

%END
\end{document}

%% file: Glossary.tex
%Al Moulla et al. (2022)

%GLOSSARY

%A:
\newacronym{ACF}{ACF}{auto-correlation function}
\newacronym{ADC}{ADC}{atmospheric dispersion corrector}

%F:
\newacronym{FWHM}{FWHM}{full-width at half-maximum}

%G:
\newacronym{GLS}{GLS}{Generalized Lomb-Scargle}

%H:
\newacronym{HELIOS}{HELIOS}{HARPS Experiment of Light Integrated Over the Sun}

%M:
\newacronym{MCMC}{MCMC}{Markov Chain Monte Carlo}

%R:
\newacronym{RMS}{RMS}{root mean square}
\newacronym{RV}{RV}{radial velocity}

%S:
\newacronym{SDO}{SDO}{Solar Dynamical Observatory}
\newacronym{SNR}{SNR}{signal-to-noise ratio}

%V:
\newacronym{VPS}{VPS}{velocity power spectrum}
\newacronym{VPSD}{VPSD}{velocity power spectral density}

%% file: Tables/Tab_01.tex
\begin{table}[t!]
    \caption{Descriptions of foreground (fg) and background (bg) likelihood parameters.}
	\begin{tabular*}{\linewidth}{l @{\extracolsep{\fill}} l l}
		\toprule
		Parameter                            & Description           & Prior                          \\
		\midrule
		$\theta_1 = \hat{c}_0$               & fg line intercept     & $\mathcal{U}$[$-0.4$, $0.4$]   \\
		$\theta_2 = c_1$                     & fg line slope         & $\mathcal{U}$[$0.075$, $0.4$]  \\
		$\theta_3 = \sigma_{\mathrm{jit}}$   & fg jitter             & $\mathcal{U}$[$0.001$, $0.08$] \\
		$\theta_4 = \hat{\mu}_{\mathrm{bg}}$ & bg mean               & $\mathcal{U}$[$0.0$, $0.8$]    \\
		$\theta_5 = \sigma_{\mathrm{bg}}$    & bg standard deviation & $\mathcal{U}$[$0.1$, $0.7$]    \\
		\bottomrule
	\end{tabular*}
	\label{Tab:1}
\end{table}

%% file: Tables/Tab_02.tex
\begin{table*}[h!]
    \caption{Best-fit coefficients and their standard deviations of fitted Lorentz and Harvey functions in Fig.~\ref{Fig:6}. Values without uncertainties, i.e. the power-law slopes of the Harvey functions and the central frequencies of the aliases, are listed without errors because they were assigned fixed values. The rotational components were constrained as harmonics, and thus the inverse central frequencies, $\nu_0^{-1}$, of the 1\textsuperscript{st} and 2\textsuperscript{nd} harmonics are exactly one half and one third, respectively, of the value for the fundamental.}
	\begin{tabular*}{\textwidth}{l @{\extracolsep{\fill}} c @{\extracolsep{\fill}} c @{\extracolsep{\fill}} c}
		\toprule
		\midrule
		\textbf{Component} & \multicolumn{3}{c}{\textbf{Coefficients}} \\
		\midrule
		& \multicolumn{3}{c}{Constants} \\
		& $C \ \left[(\mathrm{m/s})^{2}\,/\,\mathrm{Hz}\right]$ & &
		\\
		\midrule
    	Photon noise                        & $(1.39\,{\pm}\,0.27)\,{\times}\,10^{1}$ &                                          &                                          \\
		\midrule
		& \multicolumn{3}{c}{Lorentz functions} \\
		& $A_{\mathrm{L}} \ \left[(\mathrm{m/s})^{2}\,/\,\mathrm{Hz}\right]$
		& $\Gamma \ \left[\mathrm{mHz}\right]$
		& $\nu_0^{-1} \ \left[\mathrm{d}\right]$
		\\
		\midrule
		Oscillations                        & $(1.25\,{\pm}\,0.24)\,{\times}\,10^{3}$ & $(2.89\,{\pm}\,0.37)\,{\times}\,10^{-1}$ & $(3.66\,{\pm}\,0.03)\,{\times}\,10^{-3}$ \\
		Rotation fundamental                & $(5.98\,{\pm}\,1.38)\,{\times}\,10^{5}$ & $(3.41\,{\pm}\,0.71)\,{\times}\,10^{-5}$ & $(2.73\,{\pm}\,0.02)\,{\times}\,10^{1}$  \\
		Rotation 1$^{\text{st}}$ harmonic   & $(5.08\,{\pm}\,1.43)\,{\times}\,10^{5}$ & $(5.09\,{\pm}\,1.23)\,{\times}\,10^{-5}$ & $(1.36\,{\pm}\,0.01)\,{\times}\,10^{1}$  \\
		Rotation 2$^{\text{nd}}$ harmonic   & $(9.13\,{\pm}\,1.60)\,{\times}\,10^{4}$ & $(4.00\,{\pm}\,0.94)\,{\times}\,10^{-4}$ & $(9.09\,{\pm}\,0.08)\,{\times}\,10^{0}$  \\
		Alias at $\nu_0=1\,\text{day}^{-1}$ & $(1.84\,{\pm}\,0.49)\,{\times}\,10^{5}$ & $(7.70\,{\pm}\,1.75)\,{\times}\,10^{-4}$ & $1.00$                                   \\
		Alias at $\nu_0=2\,\text{day}^{-1}$ & $(1.10\,{\pm}\,1.52)\,{\times}\,10^{5}$ & $(5.17\,{\pm}\,4.92)\,{\times}\,10^{-4}$ & $0.50$                                   \\
		\midrule
		& \multicolumn{3}{c}{Harvey functions} \\
		& $A_{\mathrm{H}} \ \left[(\mathrm{m/s})^{2}\,/\,\mathrm{Hz}\right]$
		& $\tau \ \left[\mathrm{hr}\right]$
		& $\alpha$ \\
		\midrule
		Granulation                         & $(4.11\,{\pm}\,1.35)\,{\times}\,10^{2}$ & $(9.11\,{\pm}\,2.00)\,{\times}\,10^{-1}$ & $2.00$                                   \\
		Supergranulation                    & $(2.79\,{\pm}\,0.28)\,{\times}\,10^{4}$ & $(1.33\,{\pm}\,0.11)\,{\times}\,10^{1}$  & $2.00$                                   \\
		\bottomrule
	\end{tabular*}
	\label{Tab:2}
\end{table*}

%% file: Tables/Tab_03.tex
\begin{table}[t!]
    \caption{RV RMS contributions. The observed RVs correspond to the data shown in the two upper rows of Fig.~\ref{Fig:5} and the background data in Fig.~\ref{Fig:7}, and the recreated RVs correspond to the simulated contributions shown in Fig.~\ref{Fig:7}.}
	\begin{tabular*}{\linewidth}{l @{\extracolsep{\fill}} c}
		\toprule
		\midrule
		\textbf{Component} & $\bm{\mathrm{RMS} \ \left[\mathrm{m/s}\right]}$ \\
		\midrule
		\multicolumn{2}{c}{Observed RVs}           \\
		HARPS                               & 1.66 \\
		HARPS detrended with linear drift   & 1.58 \\
		HARPS detrended with 1-day mean     & 0.93 \\
		HARPS-N                             & 1.74 \\
		HARPS-N detrended with 100-day mean & 1.37 \\
		\midrule
		\multicolumn{2}{c}{Recreated RVs}          \\
		Photon noise                        & 0.24 \\
		Oscillations                        & 0.73 \\
		Rotation                            & 0.74 \\
		Granulation                         & 0.31 \\
		Supergranulation                    & 0.68 \\
		Total                               & 1.31 \\
		\bottomrule
	\end{tabular*}
	\label{Tab:3}
\end{table}